\documentclass[aps,prd,twocolumn,showpacs,eqsecnum,nofootinbib,amsmath,amssymb,amsfonts,english,jpgimgs]{revtex4}

\usepackage{babel}
\usepackage{graphicx}
\begin{document}

\title{Accelerated black holes in an anti-de~Sitter universe}

\author{Pavel Krtou\v{s}}
\email{Pavel.Krtous@mff.cuni.cz}

\affiliation{%%
  Institute of Theoretical Physics,
  Faculty of Mathematics and Physics, Charles University in Prague,\\
  V Hole\v{s}ovi\v{c}k\'{a}ch 2, 180 00 Prague 8, Czech Republic
  }

\date[Version ]{2.00; \today}

\begin{abstract}
The $C$-metric is one of few known exact solutions of 
Einstein's field equations which describes the gravitational field 
of moving sources. For a vanishing or positive cosmological constant, 
the $C$-metric represents two accelerated black holes in 
asymptotically flat or de Sitter spacetime. For a negative cosmological 
constant the structure of the spacetime is more complicated. 
Depending on the value of the acceleration, it can represent 
one black hole or a sequence of pairs of accelerated black holes in the 
spacetime with an anti-de Sitter-like infinity. 
The global structure of this spacetime is analyzed and compared 
with an empty anti-de Sitter universe. It is illustrated by 
3D conformal-like diagrams.
\end{abstract}

% 04.20.Ha Asymptotic structure
% 04.20.Jb Exact solutions

%\vspace*{0em}

\maketitle

%%%%%%%%%%%%%%%%%%%%%%%%%%%%%%%%%%%%%%%%%%%%%%%%%%%%%%%%%%%%%%%%%%%%%%
%%
%%  body
%%

\section{Introduction}
\label{sc:intro}

The $C$-metric without cosmological constant ${\Lambda}$ is a well-known solution of the Einstein(-Maxwell)
equations. It belongs to a class of spacetimes with
boost-rotational symmetry \cite{BicakSchmidt:1989} which represent
the gravitational field of uniformly accelerated sources. The $C$-metric was discovered back
in 1917 by Levi-Civita \cite{LeviCivita:1917} and Weyl \cite{Weyl:1919}, and named by Ehlers and
Kundt \cite{EhlersKundt:1962}. An understanding of the global structure
of the $C$-metric spacetime as a universe with a pair of accelerated
black holes came with the fundamental papers by Kinnersley and Walker
\cite{KinnersleyWalker:1970}, Ashtekar and Dray \cite{AshtekarDray:1981}
and Bonnor \cite{Bonnor:1982}.
Various aspects and properties of this solution were consequently studied, 
including the generalization to spinning black holes. 
References and overviews can be found, e.g., 
in Refs.~\cite{BicakSchmidt:1989,BicakPravda:1999,LetelierOliveira:2001,PravdaPravdova:2000};
for recent results see, e.g., Refs.~\cite{HongTeo:2003,HongTeo:2004,GriffithsPodolsky:2005}.

A generalization of the standard $C$-metric for nonvanishing 
cosmological constant $\Lambda$ has also been known for
a long time \cite{PlebanskiDemianski:1976,Carter:1968,Debever:1971}.
However, until recently a complete understanding of global structure 
of this solutions was missing. It was elucidated in series of papers
\cite{PodolskyGriffiths:2001,DiasLemos:2003b,KrtousPodolsky:2003}
in the case ${\Lambda>0}$, and in 
Refs.~\cite{Podolsky:2002,DiasLemos:2003a,PodolskyOrtaggioKrtous:2003}
for ${\Lambda<0}$ (cf.\ also 
Refs.~\cite{DiasLemos:2003c,Emparanetal:2000a,Chamblin:2001,Emparanetal:2000b}
for related work and discussion of special and degenerated cases).

The $C$-metric is one of few explicitly known spacetimes 
representing the gravitational field of nontrivially moving sources. Therefore,
it is interesting, for example, as a test-bed for numerical simulations.
It plays also an important role in a study of radiative properties 
of gravitational fields. Namely, in the case of nonvanishing cosmological 
constant it may provide us with an insight
into the character of radiation, which in the 
asymptotically nontrivial spacetimes is not yet well understood. 
In Refs.~\cite{KrtousPodolsky:2003,PodolskyOrtaggioKrtous:2003}
the $C$-metric spacetimes with ${\Lambda\neq0}$ were
used to investigate the directional structure of radiation.
These results were later generalized \cite{KrtousPodolsky:2004b,KrtousPodolsky:2005a}
for general spacetimes with spacelike and timelike conformal infinity.
The $C$-metric spacetimes have found also successful application to the
problem of cosmological pair creation of black holes
\cite{Emparan:1995a,Emparan:1995b,Mann:1997,BoothMann:1999,DiasLemos:2004,Dias:2004}.
In addition to spacetime with accelerated black holes, the $C$-metric can 
also describe accelerated naked singularities or, for special choice of parameters,
empty spacetime described in a coordinate system adapted to accelerated observers
\cite{PodolskyGriffiths:2001,BicakKrtous:2005,Krtous:BIAS}.

In the present work we wish to give a complete description of
the case when the $C$-metric describes black holes moving
with an acceleration in anti-de~Sitter universe. 
As was already observed in \cite{Emparanetal:2000b,DiasLemos:2003a,PodolskyOrtaggioKrtous:2003},
there are three qualitatively different cases according 
to value of the black hole acceleration ${\accl}$.
For small values of acceleration, ${\accl<1/\scl}$, 
(${\scl}$~being a length scale given by the cosmological constant, cf.\ Eq.~\eqref{scldef})
the $C$-metric describes one accelerated black hole in asymptotically
anti-de~Sitter spacetime. For large acceleration, ${\accl>1/\scl}$,
it describes a sequence of pairs of black holes. In the 
critical case ${\accl=1/\scl}$ it describes a sequence of single
accelerated black holes entering and leaving asymptotically anti-de~Sitter spacetime.
Here we concentrate on the generic situation ${\accl\neq1/\scl}$;
the critical case will be discussed separately \cite{SladekKrtous:CADSex}
(cf.\ also Refs.~\cite{Emparanetal:2000a,Chamblin:2001}).

The main goal of the work is to give a clear visual representation
of the global structure of the spacetimes. It is achieved with help
of a number of two-dimensional and three-dimensional diagrams.
Also, the relation to an empty anti-de~Sitter universe is explored.
Understanding of the anti-de~Sitter spacetime in accelerated coordinates
plays a key role in the construction of three-dimensional diagrams 
for the full $C$-metric spacetime.

The paper is organized as follows. In Sec.~\ref{sc:Cmetric} we
overview the $C$-metric solution with a negative cosmological
constant in various coordinate systems. Namely, we introduce
coordinates ${\tau,\y,\x,\ph}$, closely related to those of 
\cite{KinnersleyWalker:1970} and \cite{PlebanskiDemianski:1976},  
accelerated static coordinates ${\Tac,\Rac,\Thac,\Phac}$,
very useful for physical interpretation,
and global null coordinates ${\uG,\vG}$ essential for a study
of the global structure. In Secs.~\ref{sc:CAdSI} and \ref{sc:CAdSII}
we discuss the two qualitatively different cases of small and large 
acceleration, respectively. Finally, Sec.~\ref{sc:AdS} studies 
the weak field limit, i.e., the limit of vanishing mass and charge.
In this case the $C$-metric describes empty-anti~de-Sitter universe
in accelerated coordinates. The relation of these coordinates
to the standard cosmological coordinates is presented, 
again separately for ${\accl\lessgtr1/\scl}$.

Even more elaborated visual presentation of
the studied spacetimes, including animations and interactive three-dimensional 
diagrams, can be found in \cite{Krtous:webCADS}.
Let us also note that the on-line version of this work
includes figures in color.

\section{The $C$-metric with a negative cosmological constant}
\label{sc:Cmetric}

The  $C$-metric with a cosmological constant ${\Lambda<0}$ can be written as
\begin{equation}\label{KWmetric}
  \mtrc =
  \frac1{\accl^2(\xKW+\yKW)^2}\Bigl(
    -\FKW \,\grad\tKW^2
    +\frac1{\FKW} \,\grad\yKW^2
    +\frac1{\GKW} \,\grad\xKW^2
    +\GKW \,\grad\pKW^2
    \Bigr)\commae
\end{equation}
where $\FKW$ and $\GKW$ are polynomially dependent on 
$\yKW$ and $\xKW$ respectively,
\begin{equation}\label{KWFG}
\begin{aligned}
  \FKW &= \frac{1}{\accl^2\scl^2}-1+\yKW^2
  -2\mass\accl\,\yKW^3+\charge^2\accl^2\,\yKW^4\commae\\
  \GKW &= \mspace{64mu} 1-\xKW^2
  -2\mass\accl\,\xKW^3-\charge^2\accl^2\,\xKW^4\period
\end{aligned}
\end{equation}
Here ${\scl}$ is a length scale given by the cosmological 
constant~${\Lambda}$,
\begin{equation}\label{scldef}
  \scl = \sqrt{-\frac{3}{\Lambda}}\period
\end{equation}
The metric is a solution of the Einstein-Maxwell
equations with the electromagnetic field given by
\begin{equation}\label{KWEMF}
  \EMF = \charge\, \grad\yKW\wedge\grad\tKW\period
\end{equation}

Depending on the choice of parameters and of ranges of coordinates,
the metric \eqref{KWmetric} can describe different spacetimes.
In the physically most interesting cases, it
describes black holes uniformly accelerated in anti-de~Sitter universe.
In these cases the constants $\accl$, $\mass$, $\charge$, and $\conpar$
(such that ${\pKW\in(-\pi\conpar,\pi\conpar)}$)
characterize the acceleration, mass and charge of the black holes, and the conicity
of the $\pKW$ symmetry axis, respectively.
These parameters have to satisfy ${\mass\ge0}$, ${\charge^2<\mass^2}$, ${\accl,\,C>0}$, 
and the function ${\GKW}$ must be vanishing
for four different values of ${x}$ in the charged case (${\charge,\,\mass\neq0}$), or
for three different values in the uncharged case (${\charge=0}$, ${\mass\neq0}$). 
The coordinate ${\xKW}$ must belong to an interval around zero 
on which ${\GKW}$ is positive, and ${\yKW\in(-\xKW,\infty)}$,
cf.\ Figs.~\ref{fig:CADSI_xy} and \ref{fig:CADSII_xy}.
It follows that $0\le\GKW\le1$.
The boundary values of the allowed range of the coordinate ${\xKW}$ correspond to 
different parts of the axis of $\pKW$ symmetry separated from each other by black holes.

The spacetime described by the $C$-metric is static and axially-symmetric 
with Killing vectors $\cvil{\tKW}$ and $\cvil{\pKW}$, respectively. 
Killing horizons of the vector $\cvil{\tKW}$ are given by condition
${\FKW=0}$. They coincide with horizons of various kinds as will be described below.
Beside the Killing vectors, the geometry of spacetime
possesses one conformal Killing tensor $\tens{Q}$,
\begin{equation}\label{KillTens}
  \tens{Q} =
  \frac1{\accl^4(\xKW+\yKW)^4}\Bigl(
    \FKW \,\grad\tKW^2
    -\frac1{\FKW} \,\grad\yKW^2
    +\frac1{\GKW} \,\grad\xKW^2
    +\GKW \,\grad\pKW^2
    \Bigr) \period
\end{equation}
There exist two doubly-degenerate principal null directions
\begin{equation}\label{PNDsKW}
   \kG_1 \propto\cvil{\tKW}-\FKW\cvil{\yKW}   \comma
   \kG_2 \propto\cvil{\tKW}+\FKW\cvil{\yKW}   \commae
\end{equation}
so that the spacetime is of the Petrov type $D$.
The metric has a curvature singularity for ${\yKW\to\pm\infty}$.

The constants ${\mass}$ and ${\charge}$ parametrize the mass and charge of black holes.
Let us emphasize that they are not directly the mass or charge defined
through some invariant integral procedure. For example, the total charge 
defined by integration of the electric field over a surface around one
black hole is ${Q=\frac12\Delta\xKW\conpar\charge}$. 
It is proportional to ${\charge}$, but besides the trivial dependence
on the conicity ${\conpar}$, it depends also on the mass and the acceleration
parameters through the length ${\Delta\xKW}$ of the allowed range of the coordinate~${\xKW}$.  

The parameter ${\conpar}$ defines a range of the angular coordinate 
${\pKW}$, and thus it governs a regularity of the $\pKW$ symmetry axis.
Typically, the axis has a conical singularity which corresponds to
a string or strut. By an appropriate choice of ${\conpar}$,
a part of the axis can be made regular. However, for nonvanishing
acceleration it is not possible to achieve regularity of the whole 
axis---objects on the axis are physically responsible
for the \vague{accelerated motion} of black holes.

The constant ${\accl}$ parametrizes the acceleration of the black holes. But it is not
a simple task to define what it is the \emph{acceleration} of a black hole.
The acceleration of a test particle is defined with respect of a local inertial frame 
given by a background spacetime. However, black holes are objects which deform 
the spacetime in which they are moving; they define the notion of inertial observers, 
and they are actually dragging inertial frames with themselves. Therefore,
it is not possible to measure the acceleration of black holes with respect
to their surroundings. The motion of black holes can be partially deduced from a structure
of the whole spacetime, e.g., from a relation of black holes and asymptotically free observers,
and partially by investigating a weak field limit in which the black holes
become test particles and cease to deform the spacetime around them.
Namely, in the limit of vanishing mass and charge, the spacetime \eqref{KWmetric} reduces 
to the anti-de~Sitter universe with black holes changed into worldlines of uniformly accelerated particles.
Such a limit will be discussed in Sec.~\ref{sc:AdS}.

Depending on the value of the parameter ${\accl}$, the metric \eqref{KWmetric}
describes qualitatively different spacetimes. For ${\accl}$ smaller then 
a critical value ${1/\scl}$\, given by the cosmological constant, cf.\ Eq.~\eqref{scldef},
the metric represents asymptotically anti-de~Sitter universe
with one uniformly accelerated black hole inside.\footnote{%
As for non-accelerated black holes, it is possible to extend the spacetime
through interior of the black hole to other 
asymptotically anti-de~Sitter domain(s). However, for ${\accl<1/\scl}$, 
there is only one black hole in each of these domains.}
For ${\accl>1/\scl}$ the metric \eqref{KWmetric} describes
asymptotically anti-de~Sitter spacetime which contains
a sequence of pairs of uniformly accelerated black holes
which enter and leave the universe through its conformal infinity.\footnote{%
Again, there can be more asymptotically anti-de~Sitter domains,
each of them with the described structure.} The extremal case
${\accl=1/\scl}$ corresponds to accelerated black holes
entering and leaving the anti-de~Sitter universe, one at a time.
This extreme case will not discussed here; however, see 
Refs.~\cite{Emparanetal:2000a,Chamblin:2001,SladekKrtous:CADSex}.
 
Coordinates ${\tKW,\yKW,\xKW,\pKW}$ can be rescaled in a various way. 
We will introduce coordinates ${\tau,\y,\x,\ph}$ and closely related 
accelerated static coordinates ${\Tac,\Rac,\Thac,\Phac}$ which are appropriate for 
a discussion of the limits of weak field and of vanishing acceleration.
They will be used thoroughly in the following sections.
We will also mention coordinates ${\tPD,\yPD,\xPD,\pPD}$ (used in 
Ref.~\cite{PlebanskiDemianski:1976}) in which the global prefactor ${\accl^{-2}}$ in the
metric \eqref{KWmetric} is transformed into metric functions,
coordinates ${\tau,\om,\sm,\ph}$ adapted to the infinity, and global null coordinates
${\uG,\vG,\x,\ph}$. However, detailed transformations among these coordinates 
differs for the qualitatively different cases ${\accl\lessgtr1/\scl}$. Therefore, we 
list first only metric forms in these coordinate systems
and coordinate transformation which are general,
and we postpone specific definitions to the next sections.

The metric \eqref{KWmetric} in the coordinate systems 
${\tKW,\yKW,\xKW,\pKW}$, ${\tau,\y,\x,\ph}$ and ${\tPD,\yPD,\xPD,\pPD}$ 
has actually the same form, only with different metric functions 
(cf.\ Eqs.~\eqref{omdefI}, \eqref{FGrelI}, and \eqref{omdefII}, \eqref{FGrelII})
\begin{align}
    \mtrc & = \frac{\scl^2}{\om^2}\;\Bigl(
    -\F \,\grad\tau^2
    +\frac1{\F} \,\grad\y^2
    +\frac1{\G} \,\grad\x^2
    +\G \,\grad\ph^2
    \Bigr)\period\label{Cmetric}\\
  \mtrc & = \frac{\scl^2}{(\xPD+\yPD)^2}\Bigl(
    -\FPD \,\grad\tPD^2
    \!+\frac1{\FPD} \,\grad\yPD^2
    \!+\frac1{\GPD} \,\grad\xPD^2
    \!+\GPD \,\grad\pPD^2
    \Bigr)\!\commae\label{PDmetric}
\end{align}
Accelerated static coordinates ${\Tac,\Rac,\Thac,\Phac}$ are given by
\begin{equation}\label{acdef}
\begin{gathered}
  \Tac=\scl\tau\comma
  \Rac=\frac\scl\y\comma
  \Phac=\ph\commae\\
  \grad\Thac=\frac1{\sqrt\G}\,\grad\x\comma\quad
  \Thac=\frac\pi2\quad\text{for}\quad\x=0\period
\end{gathered}
\end{equation}
The metric takes a form 
\begin{gather}\label{Acmetric}
  \mtrc =\frac{\scl^2}{\om^2\Rac^2}\;\Bigl(
    -\Hac \,\grad\Tac^2
    +\frac1{\Hac} \,\grad\Rac^2
    +\Rac^2\bigl(\grad\Thac^2 +\G \,\grad\Phac^2\bigr)
    \Bigr)\commae\\
  \Hac=\frac1\y\,\F\smcol\label{Hacdef}
\end{gather}
see \eqref{HacI} and \eqref{HacII}.

The coordinate ${\Rac}$ is not well-defined at ${\y=0}$. It
is a coordinate singularity which can be avoided by using the
coordinate~${\y}$. However, near the black hole, the coordinate ${\Rac}$ 
has a more direct physical meaning---it is the radial coordinate measured by area, 
at least in the conformally related geometry.
Because ${\y}$ can be negative, ${\Rac}$ can take also negative values.
However, it happens only far away from the black holes or in spacetime domains
in which ${\Rac}$ changes into a time coordinate.

The coordinate ${\x}$ is given by ${\x=-\xKW}$ 
(cf.\ Eqs.~\eqref{coordefI} and \eqref{coordefII}), so we can use what was said
about range of definition of ${\xKW}$. Let ${[\xb,\xf]}$ be the interval of 
allowed values of ${\x}$, i.e., the interval where
${\G}$ is positive and ${\xb<0<\xf}$. The value ${\xf}$ corresponds to the axis
of ${\ph}$ symmetry (since ${\G=0}$ at ${\x=\xf}$)  
pointing out of the black hole in the forward direction of the motion.\footnote{%
By the direction of motion we mean the direction from which the black hole is
pulled by the cosmic string or toward which it is pushed by the strut. 
In the weak field limit it is the direction of the acceleration.
}
The value ${\xb}$ corresponds to the axis (again, ${\G\vert_{\x=\xb}=0}$) 
going in the opposite (backward) direction.
Integrating ${1/\sqrt\G}$ in \eqref{acdef}, we find that the longitudinal angular 
coordinate ${\Thac}$ belongs into an interval ${[\Thac_\baxis,\Thac_\faxis]}$
which, in general, differs from ${[0,\pi]}$.

If we use the conformal prefactor in the metric \eqref{Cmetric}
as a coordinate, and if we find a complementary coordinate ${\sm}$
such that the metric is diagonal (see \eqref{osdifI} and \eqref{osdifII}),
we get
\begin{equation}\label{osmetric}
  \mtrc = \frac{\scl^2}{\om^2}\;\Bigl(
    -\F \,\grad\tau^2
    +\frac1{\E}\,\bigl(\grad\om^2
    +\F\G \,\grad\sm^2\bigr)
    +\,\G \,\grad\ph^2
    \Bigr)\commae
\end{equation}
This coordinate system is well adapted to the infinity ${\scri}$, since ${\scri}$
is given by ${\om=0}$.

Finally, for discussion of global structure of the spacetime it is
useful to introduce global null coordinates\footnote{%
Notice the difference between $\vG$~(v) and
$\y$~(upsilon). It should be always clear from the context 
if we speak about null~${\vG}$ or radial~${\y}$.}
${\uG,\vG,\x,\ph}$.
We start with the \vague{tortoise} coordinate ${\yt}$
\begin{equation}\label{ytort}
  \grad\yt=\frac1\F\,\grad\y
\end{equation}   
It expands each of the intervals between successive zeros of ${\F}$ to the whole real line.
Next we define null coordinates
${\uB,\vB}$
\begin{equation}\label{uvBlock}
  \uB=\yt+\tau\comma\vB=\yt-\tau\period
\end{equation} 
These coordinates cover distinct domains of the spacetime which are separated 
from each other by horizons, i.e., by null surfaces ${\F=0}$. The coordinates can be extended 
across a chosen horizon with help of global coordinates ${\uG,\vG}$:
\begin{equation}\label{uvGlob}
\begin{gathered}
  \tan\frac{\uG}{2}=(-1)^m\exp\frac{\uB}{2\abs{\GNcoef}}\commae\\
  \tan\frac{\vG}{2}=(-1)^n\exp\frac{\vB}{2\abs{\GNcoef}}\period
\end{gathered}
\end{equation}
Integers ${m,n}$ label the domains; see Figs.~\ref{fig:CADSI_cd},
\ref{fig:CADSII_cd} and \ref{fig:ADSII_cd} below. ${\GNcoef}$ is a real constant.
The metric reads\footnote{%
${\grad\uG\stp\grad\vG=\grad\uG\,\grad\vG+\grad\vG\,\grad\uG}$ is a symmetric tensor product,
which is usually loosely written as ${2\grad\uG\grad\vG}$.}
\begin{equation}\label{GNmetric}
  \mtrc = \frac{\scl^2}{\om^2}\;\Bigl(
    \frac{2\,\GNcoef^2\F}{\sin\uG\sin\vG}\,
    \grad\uG\stp\grad\vG
    +\frac1\G\,\grad\x^2+\G\,\grad\ph^2\Bigr)\period
\end{equation}
For a suitable choice of the constant ${\GNcoef}$ the metric coefficients
turn to be smooth and nondegenerate  as functions of coordinates 
${\uG,\vG}$ across a chosen horizon. 
For such a choice we require that the coordinate map ${\uG,\vG,\x,\ph}$ on a neighborhood
of that horizon belongs to the differential atlas of the manifold.
The metric is thus smoothly extended across the chosen horizon.

\section{A single accelerated black hole}
\label{sc:CAdSI}

\subsection{Coordinate systems}

We start a specific discussion with the simpler case 
\begin{equation}\label{acclI}
\accl<\frac1\scl\period
\end{equation} 
The coordinates ${\tau,\y,\x,\ph}$ and ${\tPD,\yPD,\xPD,\pPD}$ are in this case defined by
\begin{equation}\label{coordefI}
\begin{aligned}
  \tau&=\cos\aca\;\tPD&&=\cot\aca\;\tKW\commae\\
  \y&=\frac1{\cos\aca}\;\yPD&&=\tan\aca\;\yKW\commae\\
  \x&=-\frac1{\sin\aca}\;\xPD\mspace{-12mu}&&=-\xKW\commae\\
  \ph&=\sin\aca\;\pPD&&=\pKW\commae
\end{aligned}
\end{equation}
where ${\aca\in[0,\frac\pi2)}$ is a parameter characterizing the acceleration,
\begin{equation}\label{acadef}
  \accl=\frac1\scl\,\sin\aca\period
\end{equation}
Its geometrical meaning in the weak field limit will be discussed in Sec.~\ref{sc:AdS}.
The metric functions in \eqref{Cmetric} and \eqref{PDmetric} are given by
\begin{gather}\label{CFGI}
\begin{aligned}
  \F &= 1+\y^2-2\,\frac\mass\scl\,\cos\aca\;\y^3+\frac{\charge^2}{\scl^2}\,\cos^2\aca\;\y^4\commae\\
  \G &= 1-\x^2+2\,\frac\mass\scl\,\sin\aca\;\x^3-\frac{\charge^2}{\scl^2}\,\sin^2\aca\;\x^4\commae
\end{aligned}\\
  \om=\y\,\cos\aca-\x\,\sin\aca\commae\label{omdefI}
\end{gather}
and
\begin{equation}\label{PDFGI}
\begin{aligned}
  \FPD&= \cos^2\aca+\yPD^2-2\,\frac\mass\scl\,\yPD^3+\frac{\charge^2}{\scl^2}\,\yPD^4\commae\\
  \GPD&= \sin^2\aca-\xPD^2-2\,\frac\mass\scl\,\xPD^3-\frac{\charge^2}{\scl^2}\,\xPD^4\period
\end{aligned}
\end{equation}
They are related by
\begin{equation}\label{FGrelI}
\begin{aligned}
  \F &\!=\! \cos^{\!-2}\!\aca\;\FPD\!=\!\tan^2\!\aca\;\FKW
  \!=\!1\!-{\textstyle\frac{\scl^2}{\cos^2\!\aca}}\Sfc\bigl({\textstyle\frac{\cos\aca}\scl\y})\commae\\
  \G &\!=\! \sin^{\!-2}\!\aca\;\GPD\!=\mspace{54mu}\GKW
  \!=\!1\!+\!{\textstyle\frac{\scl^2}{\sin^2\!\aca}}\Sfc\bigl({\textstyle\frac{\sin\aca}\scl\,\x})\commae
\end{aligned}
\end{equation}
where ${S(w)}$ is a simple polynomial
\begin{equation}\label{Sdef}
  \Sfc(w)=-w^2(1-2\mass w +\charge^2 w^2)\period
\end{equation}
The functions ${\Hac}$ is (cf.\ Eq.~\eqref{Hacdef})
\begin{equation}\label{HacI}
  \Hac=1+\frac{\Rac^2}{\scl^2}-\cos\aca\;\frac{2\mass}{\Rac}+\cos^2\aca\;\frac{\charge^2}{\Rac^2}\period
\end{equation}

The coordinate ${\om}$ was already defined in Eq.~\eqref{omdefI}. The complementary
orthogonal coordinate ${\sm}$ can be, in general, given simply only in differential form\footnote{%
The relations are integrable sice ${\F}$ depends only on ${\y}$ and ${\G}$ on~${\x}$.}
\begin{equation}\label{osdifI}
\begin{gathered}
  \grad\sm=\frac{\sin\aca}{\F}\,\grad\y+\frac{\cos\aca}{\G}\,\grad\x\commae\\
  \grad\om=-\cos\aca\,\grad\y+\sin\aca\,\grad\x\period
\end{gathered}
\end{equation}
(Here we included also the gradient of ${\om}$ for completeness.)
The metric function ${\E}$ is given by 
\begin{equation}\label{EdefI}
  \E=\F\cos^2\aca+\G\sin^2\aca\period
\end{equation}
At infinity, ${\om=0}$ and ${\E=1}$.

\subsection{Global structure}

Now we are prepared to discuss the global structure of the spacetime in more details.
We start inspecting the metric in the accelerated static coordinates \eqref{Acmetric}
with ${\Hac}$ given by \eqref{HacI}. It has a familiar form---if 
we ignore prefactor ${\scl^2/(\om\Rac)^2}$ 
we get the metric of a nonaccelerated black hole
in anti-de~Sitter universe in standard static coordinates---except for a different 
range of ${\Thac}$ and except for ${\G}$
instead of ${\sin^2\Thac}$ in front of the ${\grad\Phac^2}$ term. 
Fortunately, ${\sqrt\G}$ on the allowed range of ${\Thac}$ resembles ${\sin\Thac}$,
and the difference does not affect qualitative properties of the geometry.\footnote{%
Let us mention that for ${\accl=0}$, i.e., for ${\aca=0}$, the metric \eqref{Acmetric}
becomes exactly the Reissner-Nordstr\"om--anti-de~Sitter solution with
${\x=-\cos\Thac}$, ${\G=\sin^2\Thac}$, and 
${\Hac=1+\frac{\Rac^2}{\scl^2}-\frac{2\mass}\Rac+\frac{\charge^2}{\Rac^2}}$.} 
The conformal prefactor ${\scl^2/(\om\Rac)^2}$ does not change the causal structure 
of the black hole. It justifies our claim that the spacetime contains a black hole.
It also gives the interpretation for the coordinates---the accelerated static
coordinates are centered around the hole, with ${\Rac}$ being a radial coordinate, 
and ${\Thac}$ and ${\Phac}$ longitudinal and latitudinal angular coordinates. 
${\Tac}$ is a time coordinate of external observers staying at a constant distance above 
the horizon of the black hole.
The coordinates ${\tau,\y,\x,\ph}$ are only a different parametrization
of the time, radial and angular directions.

However, the prefactor ${\scl^2/(\om\Rac)^2}$ in \eqref{Acmetric} 
changes the \vague{position} of the infinity---the conformal infinity 
${\scri}$ is localized at ${\om=0}$, i.e., at
\begin{equation}\label{scriI}
  \y = \tan\aca\;\x\period
\end{equation}
It means that the radial position of the infinity depends on the direction ${\x}$.
This corresponds to the fact that the black hole is not in a symmetrical position 
with respect to the asymptotically anti-de~Sitter universe. Nevertheless,
it is in equilibrium---the cosmological compression of anti-de~Sitter spacetime
(which would push a test body toward any chosen center of the universe) 
is compensated by a string (or strut) on the axis which keeps the black hole in
a static nonsymmetric position with respect to the infinity. 
We can thus say that the black hole is moving with uniform acceleration
equal to the cosmological compression, 
despite the fact that it cannot be measured locally. 
Remember that in anti-de~Sitter universe  
a static observer which stays at a fixed spatial position in the spacetime eternally 
feels the cosmological deceleration
of a constant magnitude from the range ${[0,1/\scl)}$, depending on his position. 
This corresponds to the assumption \eqref{acclI}. 
As we will see in a moment, we are dealing with \emph{one} black hole
which stays \emph{eternally} in equilibrium in asymptotically anti-de~Sitter spacetime.  

\begin{figure}
  \includegraphics{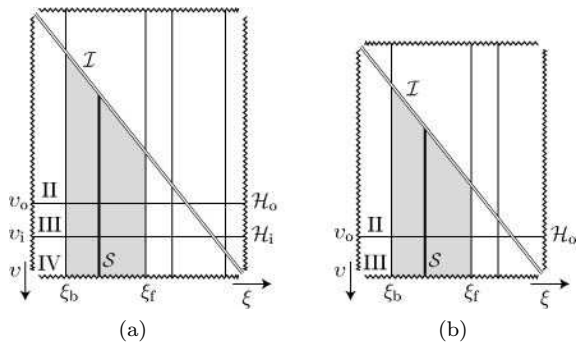}\\
  {\footnotesize\hspace*{0pt}(a)\hspace{110pt}(b)\hspace*{0pt}}
  \caption{\label{fig:CADSI_xy}%
The diagrams of the allowed range of coordinates ${\y}$ and ${\x}$ 
(the shaded region) in the case of small acceleration ${\accl<1/\scl}$. 
Diagram (a) is applicable in the charged case, 
(b) is valid for ${\mass\neq0}$, ${\charge=0}$. 
${\xb}$ and ${\xf}$ are zeros
of the metric function ${\G}$ closest to ${\x=0}$. These
values correspond to the axis of ${\ph}$ symmetry.
The diagonal double-line represents the infinity, cf.\ Eq.~\eqref{scriI}. 
The bottom zig-zag line is the singularity at ${\y=\infty}$.
${\yo}$ and ${\yi}$ are zeros of the metric function ${\F}$. 
They define the outer and inner black hole horizons.
They separate the allowed range of coordinates into 
regions II, III, and IV. These regions
corresponds to different domains in spacetime, each of them covered by 
its own coordinates ${\tau,\y,\x,\ph}$. These coordinate
systems cannot be smoothly extended over the horizon. 
Coordinates smooth across the horizon are used in Fig.~\ref{fig:CADSI_cd},
where sections ${\x=\text{constant}}$ are depicted. 
Such a section is represented in the diagrams above by the vertical thick line.
}
\end{figure}
\begin{figure}
  \includegraphics{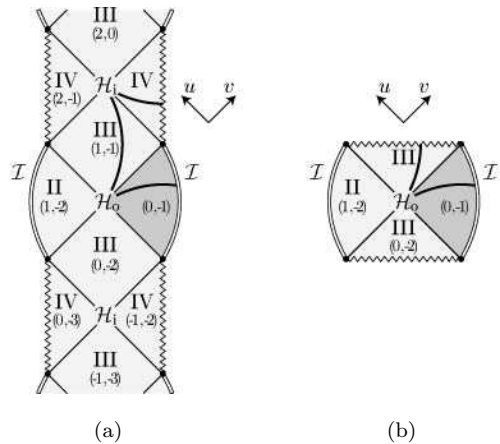}\\[6pt]
  {\footnotesize\hspace*{0pt}(a)\hspace{100pt}(b)\hspace*{0pt}}
  \caption{\label{fig:CADSI_cd}%
The conformal diagrams of the sections ${\x,\ph=\text{constant}}$
for (a) charged and (b) uncharged $C$-metric
with the acceleration ${\accl<1/\scl}$.
These diagrams are based on null coordinates ${\uG,\vG}$ which grow in diagonal directions.
Integers ${(m,n)}$ in the diagrams, identifying different spacetime domains,
are those from definition \eqref{uvGlob}.
Double-lines represent conformal infinity ${\scri}$ (cf.\ Eq.~\eqref{scriI}), 
zig-zag lines the singularity at ${\y=+\infty}$, and thin diagonal lines the outer
and inner black hole horizons ${\y=\yo}$ and ${\y=\yi}$, respectively.
We can recognize familiar structure of the interior of Reissner-Nordstr\"om
or Schwarzschild black holes, respectively (domains III and IV). 
The exterior of the black holes 
is, however, asymptotically different---it has the asymptotics of anti-de~Sitter
universe. The whole spacetime consists of more exterior domains II
which are connected (not necessary causally) with each other through the black holes.
A more detailed diagram of a typical domain outside of the black hole 
(a darker area indicated above) can be found in Fig.~\ref{fig:CADSI_3D}b.   
The thick line corresponds to a section ${\tau=\text{constant}}$ which
is discussed in Fig.~\ref{fig:CADSI_xy}.
}
\end{figure}

We already said that zeros of ${\F}$ correspond to Killing horizons of the Killing vector 
${\cvil{\tau}}$. Inspecting properties of the polynomial ${\Sfc(w)}$, we find that
${\F=0}$ for two values ${\y=\yo,\yi\;}$ (${\yo<\yi}$) in the charged case (${\charge,\mass\ne0}$),
and for just one value ${\y=\yo}$ if ${\charge=0}$, ${\mass\ne0}$. The null surface
${\y=\yo}$ corresponds to the outer black hole horizon, and ${\y=\yi}$ corresponds to the inner black hole horizon. 

Allowed ranges of coordinates ${\y,\x}$ are shown in Fig.~\ref{fig:CADSI_xy}.
Boundary \vague{zig-zag} lines corresponds to the curvature singularity at ${\y,\x\to\pm\infty}$.
The horizons separate the allowed range into qualitatively different regions II, III,
and IV. Region II describes the asymptotically anti-de~Sitter 
domain outside of the black hole, and regions III and IV correspond to 
the interior of the black hole. 

The coordinate systems ${\tau,\y,\x,\ph}$ or ${\Tac,\Rac,\Thac,\Phac}$ are defined
in each of the regions II, III, IV; however, they are singular at the horizons. 
To extend the spacetime through the horizons, global null coordinates ${\uG,\vG,\x,\ph}$
can be used. It turns out that the global manifold contains more domains 
of the type II, III, IV, labeled by integers ${m,n}$; see Eq.~\eqref{uvGlob}. 
From the domain II outside the outer black hole horizon, the spacetime continues
into two domains III inside the black hole. These are connected to other 
asymptotically anti-de~Sitter domains II (behind the Einstein-Rosen bridge through
the black hole), and, in the charged case, to domains IV behind inner black hole horizons. 
Each of these domains is covered by its own coordinate system 
${\tau,\y,\x,\ph}$. This global structure is well illustrated 
in two-dimensional conformal diagrams  
of ${\x,\ph=\text{constant}}$ sections; see Fig.~\ref{fig:CADSI_cd}.

\begin{figure}
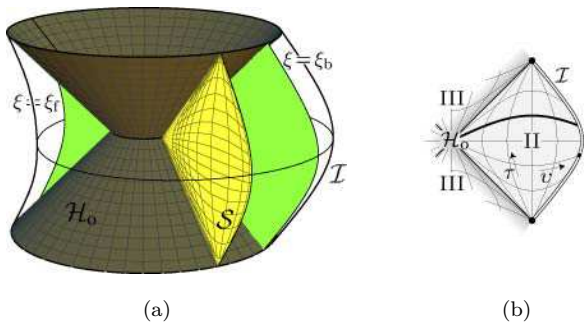

  \begin{minipage}[c]{140pt}\centering\includegraphics{\imgver fig_CAdSI_3Dcone}\end{minipage}
  \begin{minipage}[c]{100pt}\centering\includegraphics{\imgver fig_CAdSI_2D}\end{minipage}\\[9pt]
  {\footnotesize\hspace*{18pt}(a)\hspace{125pt}(b)\hspace*{0pt}}
  \caption{\label{fig:CADSI_3D}%
(a) Three-dimensional representation of the
\emph{exterior} of the black hole accelerated in anti-de~Sitter universe with
acceleration smaller then ${1/\scl}$. The dark surface represents the outer black hole horizon ${\horo}$,
and the boundary of the diagram corresponds to the conformal infinity ${\scri}$.
Embeddings of a typical section ${\x=\text{constant}}$ (section ${\sect{}}$) and of the axis ${\x=\xf,\xb}$ are shown.
The nonsymmetric shape of the infinity reflects the fact that the coordinate system used
is centered around the black hole which is moving with acceleration with respect to the infinity.
(b) Two-dimensional conformal diagram of ${\x=\text{constant}}$ section. 
Only the exterior of the black hole is shown (compare with Fig.~\ref{fig:CADSI_cd}). 
This part of the conformal diagram corresponds exactly to the section ${\x=\text{constant}}$ 
indicated in the diagram on the left.}
\end{figure}

As already mentioned, the inner structure of the black hole is qualitatively 
the same as the structure of the interior of the standard Schwarzschild or Reissner-Nordstr\"om black holes.
Therefore we focus mainly on the exterior of the black hole. 
A more detailed conformal diagram of the domain outside of the outer horizon
can be found in Fig.~\ref{fig:CADSI_3D}b. The position of the infinity in the diagrams
for various values of ${\x}$ changes according to \eqref{scriI}.
We can glue sheets of different ${\x}$ together 
into three-dimensional diagram in Fig.~\ref{fig:CADSI_3D}a,
where only the coordinate ${\ph}$ is suppressed. 
The \vague{gluing} is done using an intuition that
${\x}$ is a \vague{deformed cosine} of longitudinal angle and that
${\y}$ parametrizes the radial direction. The three-dimensional diagram 
in Fig.~\ref{fig:CADSI_3D}a is thus obtained
by a rotation of the conformal diagram in Fig.~\ref{fig:CADSI_3D}b.

\begin{figure}
  \includegraphics{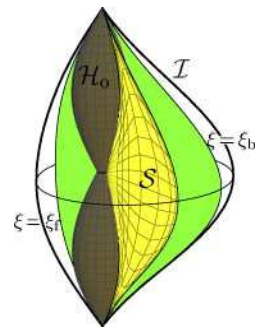}\\
  \caption{\label{fig:CADSI_3Ddrop}%
Another three-dimensional representation
of the exterior of the accelerated black hole with ${\accl<1/\scl}$. 
The outer black hole horizon of a conical shape from Fig.~\ref{fig:CADSI_3D}a
is here deformed to the surface of a shape of two joined drops. 
The black hole is thus represented as a localized object. Such a representation 
is useful for a study of the weak field limit when the black hole changes into
the worldline of a point particle.
}
\end{figure}

The outer black hole horizon has a form of two cone-like surfaces joined 
in the neck of the black hole. The conical shape suggests that
horizon is a null surface with null generators originating from the neck. Of course,
the three-dimensional diagram does not have the nice feature of the two-dimensional 
conformal diagrams that each line with angle ${\pi/4}$ from the vertical is null; however,
for Fig.~\ref{fig:CADSI_3D}a 
this feature still holds for lines in radial planes,
i.e., it holds for generators of the black hole horizon.

In the weak field limit the black hole changes into a test particle. For such a 
transformation the diagram in Fig.~\ref{fig:CADSI_3D}a is not very intuitive---the black hole
is represented there as an \vague{extended} object, and the qualitative shape of the horizon
does not change with varying mass and charge. For this reason it is useful to draw
another diagram in which the black hole horizon is deformed into
a shape composed of two drop-like surfaces, see Fig.~\ref{fig:CADSI_3Ddrop}. 
The conical form of the horizon from Fig.~\ref{fig:CADSI_3D}a
is \vague{squeezed} into more localized form, which in the limit of vanishing 
mass and charge shrinks into a worldline of the particle---cf.\ 
Fig.~\ref{fig:ADSI_3D}b in Sec.~\ref{sc:AdS}. 

\begin{figure}[b]
  \includegraphics{\imgver fig_CAdSII_xy}\\
  {\footnotesize\hspace*{0pt}(a)\hspace{105pt}(b)\hspace*{0pt}}
  \caption{\label{fig:CADSII_xy}%
Diagrams analogous to Fig.~\ref{fig:CADSI_xy}
in the case ${\accl>1/\scl}$.
The allowed range of coordinates ${\y,\x}$ (shaded area) is again
restricted by the infinity (diagonal double-line), by the axis (vertical border lines),
and by the singularity (zig-zag line). Additionally to outer
and inner black hole horizons, acceleration and cosmological
horizons (at ${\y=\ya}$ and ${\y=\yc}$) are also present. Horizons
divide the allowed range into regions O-IV which corresponds
to qualitatively different domains of spacetime; cf.\ Fig.~\ref{fig:CADSII_cd}.
Different sections ${\x=\text{constant}}$ cross different number of horizons.
Typical representives ${\sect{A}}$, ${\sect{B}}$, and ${\sect{C}}$
of these sections are indicated by thick vertical lines. They correspond
to different shapes of the conformal diagrams in 
Fig.~\ref{fig:CADSII_cd}.
}
\end{figure}

\section{Pairs of accelerated black holes}
\label{sc:CAdSII}

\begin{figure*}
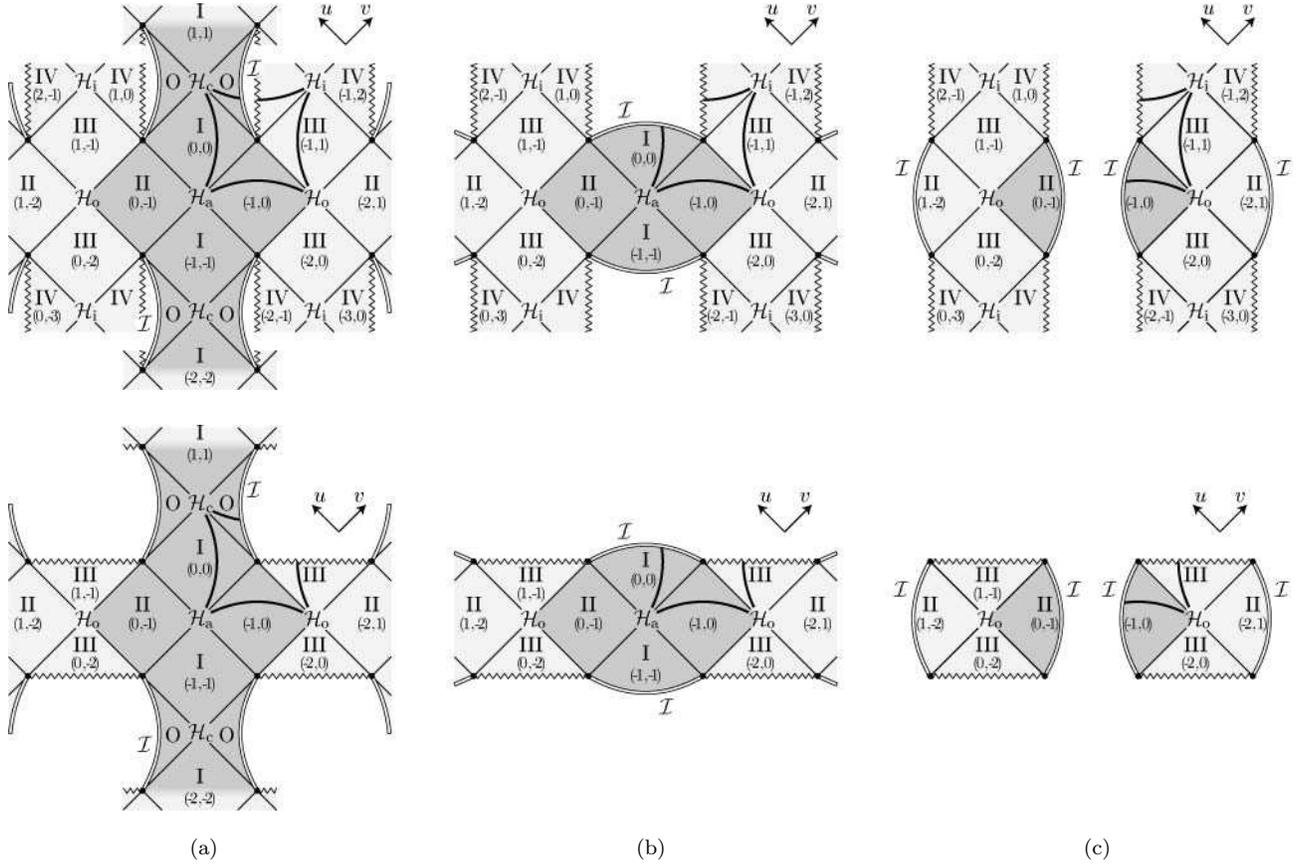

  \includegraphics{\imgver fig_CAdSII_cd_me}\\[12pt]
  \includegraphics{\imgver fig_CAdSII_cd_m}\\[6pt]
  {\footnotesize\hspace*{0pt}(a)\hspace{158pt}(b)\hspace{158pt}(c)\hspace*{0pt}}
  \caption{\label{fig:CADSII_cd}%
The conformal diagrams of the sections ${\x,\ph=\text{constant}}$ for ${\accl>1/\scl}$.
The top diagrams are valid for ${\mass,\charge\neq0}$, the bottom ones are for the uncharged case.
The diagrams are based on coordinates ${\uG,\vG}$. Integers ${(m,n)}$ from 
definition \eqref{uvGlob} identify different domains of the spacetime.
Analogously to Fig.~\ref{fig:CADSI_cd}, double lines represent the infinity, zig-zag lines the 
singularity, and diagonal lines the horizons. Domains O-II correspond to the exterior 
of black holes and domains III and IV to interiors of black holes. The interior has
a similar causal structure to that of  unaccelerated black holes. 
The spacetime contains more asymptotic domains, one of which is indicated by dark shading.
The description below is from a point of view of this domain. 
Three different shapes of the diagrams correspond to the sections with different 
value of the coordinate ${\x}$. On the left, section ${\sect{A}}$
is spanned between two black holes which are moving with respect to each other along a common axis.
It is also spanned between different pairs of such black holes through the domains O and I.
In the middle, section ${\sect{B}}$ is spanned only between two black holes.
It does not continue to other pair of black hole because it 
intersects the conformal infinity in spacelike lines located inside domains I.
The section ${\sect{C}}$, depicted on the right, goes from each black hole 
directly into infinity---it does not connect different black holes
through the exterior domains. These three sections correspond to
thick vertical lines in Fig.~\ref{fig:CADSII_xy}. Thick lines in the
diagrams above represent the section ${\tau=\text{constant}}$, i.e.,
exactly the section discussed in Fig.~\ref{fig:CADSII_xy}.
The embedding of these two-dimensional diagrams into spacetime
is shown in Figs.~\ref{fig:CADSII_3DA}--\ref{fig:CADSII_3DC}.
More detailed two-dimensional diagrams of the exterior of
black holes (the dark area above) are also presented there.  
}
\end{figure*}

\subsection{Coordinate systems}

Next we turn to the discussion of the more intricate case of the acceleration bigger than the critical one, 
\begin{equation}\label{acclII}
\accl>1/\scl\period
\end{equation} 

\begin{figure}[t]
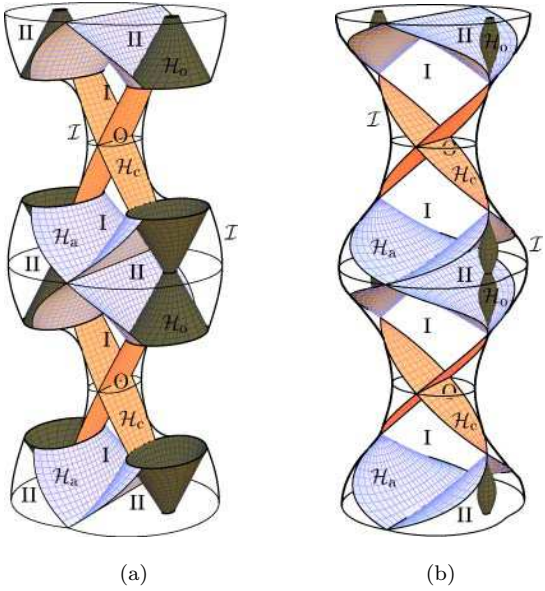

  \includegraphics{\imgver fig_CAdSII_3Dcone}\qquad\qquad
  \includegraphics{\imgver fig_CAdSII_3Ddrop}\\[6pt]
  {\footnotesize\hspace*{10pt}(a)\hspace{105pt}(b)\hspace*{0pt}}
  \caption{\label{fig:CADSII_3D}%
Three-dimensional visualizations of the \emph{exterior} of black holes
which are moving with acceleration parameter ${\accl>1/\scl}$
in asymptotically anti-de~Sitter universe. 
The diagrams show a compactified picture
of the whole universe---borders of the diagrams correspond to 
the conformal infinity. Diagram (a) is obtained by 
gluing together two-dimensional diagrams from Fig.~\ref{fig:CADSII_cd}.
Black hole outer horizons ${\horo}$ are represented by dark surfaces of a conical shape
which indicates the null character of these surfaces. 
In the alternative representation (b), the 
black hole outer horizons are squeezed into drop-like shapes. Such a
representation shows the black hole as a localized object
and it is useful in the weak field limit when the black hole
changes to a point-like particle---compare with Fig.~\ref{fig:ADSII_3D}b.
The universe represents a sequence of pairs of black holes which
repeatedly enter and leave the universe through its timelike 
infinity---the diagrams should continue periodically in the vertical direction. 
Black holes of each pair are causally separated by the acceleration horizon ${\hora}$;
consequent pairs of black holes are separated by cosmological horizons ${\horc}$. 
These are null surfaces---light cones of the \vague{entry points} of black
holes into the spacetime. Embedding of different types of two-dimensional
conformal diagrams into the three-dimensional one is 
depicted in Figs.~\ref{fig:CADSII_3DA}--\ref{fig:CADSII_3DC}.
}
\end{figure}

First, the coordinates ${\tPD,\yPD,\xPD,\pPD}$ and ${\tau,\x,\y,\ph}$ can be defined in
an analogous way as in the previous section:
\begin{equation}\label{coordefII}
\begin{aligned}
  \tau&= \sh\acp\;\tPD        &&= \tanh\acp\;\,\tKW\commae\\
  \y&=\frac1{\sh\acp}\,\yPD   &&=\coth\acp\;\,\yKW\commae\\
  \x&=-\frac1{\ch\acp}\,\xPD\mspace{-10mu}  &&=-\xKW\commae\\
  \ph&=\ch\acp\;\pPD          &&=\pKW\period
\end{aligned}
\end{equation}
Ranges of the coordinates ${\y,\x}$ are indicated in Fig.~\ref{fig:CADSII_xy}.
The acceleration is parametrized by the parameter ${\acp\in\realn^+}$,
\begin{equation}\label{acpdef}
  \accl=\frac1\scl\cosh\acp\period
\end{equation}
The metric functions in \eqref{Cmetric} and \eqref{PDmetric} are
\begin{gather}
\begin{aligned}\label{CFGII}
  -\F&= 1-\y^2+2\,\frac\mass\scl\,\sh\acp\;\y^3-\frac{\charge^2}{\scl^2}\,\sh^2\acp\;\y^4\commae\\
  \G &= 1-\x^2+2\,\frac\mass\scl\,\ch\acp\;\x^3-\frac{\charge^2}{\scl^2}\,\ch^2\acp\;\x^4\commae
\end{aligned}\\
  \om=\y\,\sh\acp-\x\,\ch\acp\commae\label{omdefII}
\end{gather}
and
\begin{equation}\label{PDGII}
\begin{aligned}
  -\FPD&= \sh^2\acp-\yPD^2+2\,\frac\mass\scl\,\yPD^3-\frac{\charge^2}{\scl^2}\,\yPD^4\commae\\
  \GPD &= \ch^2\acp-\xPD^2+2\,\frac\mass\scl\,\xPD^3-\frac{\charge^2}{\scl^2}\,\xPD^4\period
\end{aligned}
\end{equation}
They are again related to the polynomial \eqref{Sdef}
\begin{equation}\label{FGrelII}
\begin{aligned}
  -\F &\!=\! -\sh^{\!-2}\!\aca\;\FPD\!=-\!\coth^2\!\aca\;\FKW
  \!=\!1\!+{\textstyle\frac{\scl^2}{\sh^2\!\acp}}\Sfc\bigl({\textstyle\frac{\sh\acp}\scl\y})\commae\\
  \G &\!=\! \spcm\ch^{\!-2}\!\acp\;\GPD\!=\mspace{76mu}\GKW
  \!=\!1\!+\!{\textstyle\frac{\scl^2}{\ch^2\!\acp}}\Sfc\bigl({\textstyle\frac{\ch\acp}\scl\,\x})\period
\end{aligned}
\end{equation}
For the metric function ${\Hac}$, given by Eq.~\eqref{Hacdef}, we obtain 
\begin{equation}\label{HacII}
  \Hac=1-\frac{\Rac^2}{\scl^2}-\sh\acp\;\frac{2\mass}{\Rac}+\sh^2\acp\;\frac{\charge^2}{\Rac^2}\period
\end{equation}

Differential relations for the coordinates ${\sm}$ and ${\om}$ are
\begin{equation}\label{osdifII}
\begin{gathered}
  \grad\sm=\frac{\ch\acp}{\F}\,\grad\y+\frac{\sh\acp}{\G}\,\grad\x\commae\\
  \grad\om=-\sh\acp\,\grad\y+\ch\acp\,\grad\x\commae
\end{gathered}
\end{equation}
and the metric function ${\E}$ takes the form
\begin{equation}\label{EdefII}
  \E=\F\sh^2\acp+\G\ch^2\acp\period
\end{equation}

\subsection{Global structure}

\begin{figure}[t]
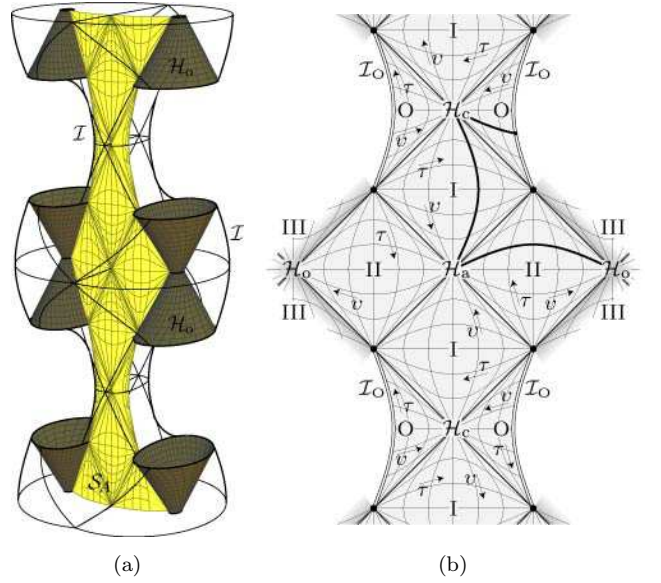

  \begin{minipage}[c]{90pt}\centering\includegraphics{\imgver fig_CAdSII_3DA}\end{minipage}
  \begin{minipage}[c]{150pt}\centering\includegraphics{\imgver fig_CAdSII_2DA}\end{minipage}\\[6pt]
  {\footnotesize\hspace*{0pt}(a)\hspace{112pt}(b)\hspace*{30pt}}
  \caption{\label{fig:CADSII_3DA}%
(a) Embedding of section ${\sect{A}}$
(cf.\ Figs.~\ref{fig:CADSII_xy} and \ref{fig:CADSII_cd}a) into 
a three-dimensional representation of the $C$-metric spacetime. 
(b) The part of the two-dimensional conformal diagram of ${\sect{A}}$ representing 
the exterior of the black holes (corresponds to the dark
area in Fig.~\ref{fig:CADSII_cd}a).
}
\end{figure}

As in the previous case, we start with a discussion of the 
metric in accelerated static coordinates, Eq.~\eqref{Acmetric}.
Near the outer and inner horizon (the smallest two zeros of ${\Hac}$), 
the metric function \eqref{HacII} has a similar behavior 
as the function \eqref{HacI}. It means that we deal again with
a black hole, and near (or inside of) the black hole we can apply 
the previous discussion. Namely, ${\Tac}$ is again
a time coordinate for observers staying outside black hole,
${\Rac}$ is a radial coordinate, and ${\Thac,\Phac}$ are spherical-like angular coordinates.
Similar interpretation hold for the coordinates ${\tau,\y,\x,\ph}$.
However, for ${\accl>1/\scl}$ the metric function ${\Hac}$ 
(or, equivalently, ${\F}$, cf.\ Eq.~\eqref{Hacdef}) 
has two additional zeros for ${\Rac=\Rac_\ahor,\Rac_\chor}$
(${\y=\ya,\yc}$, respectively), which correspond to
acceleration and cosmological horizons. 
It means that we have to expect a more
complicated structure of spacetime outside the black hole.

\begin{figure}[t]
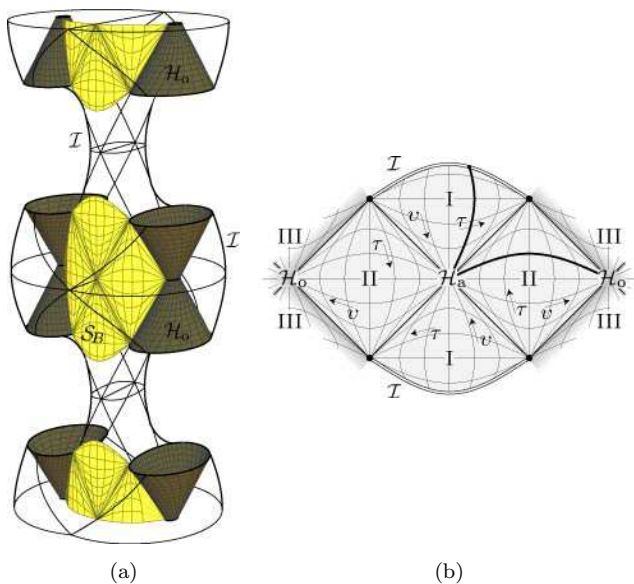

  \begin{minipage}[c]{90pt}\centering\includegraphics{\imgver fig_CAdSII_3DB}\end{minipage}
  \begin{minipage}[c]{150pt}\centering\includegraphics{\imgver fig_CAdSII_2DB}\end{minipage}\\[6pt]
  {\footnotesize\hspace*{0pt}(a)\hspace{112pt}(b)\hspace*{30pt}}
  \caption{\label{fig:CADSII_3DB}%
(a) Embedding of section ${\sect{B}}$
(cf.\ Figs.~\ref{fig:CADSII_xy} and \ref{fig:CADSII_cd}b) into 
a three-dimensional picture of spacetime. 
(b) The corresponding
part of the two-dimensional conformal diagram of ${\sect{B}}$
(cf.\ the dark area in Fig.~\ref{fig:CADSII_cd}b).  
}
\end{figure}

Indeed, from the ${\x\textdash\y}$ diagram in Fig.~\ref{fig:CADSII_xy}
we see that new zeros divide the allowed range of coordinates into more 
regions O, I, II, III, and IV.
An exact way how these domains can be reached through the horizons
can be seen from the conformal diagrams of the sections ${\x,\ph=\text{constant}}$.
However, in Fig.~\ref{fig:CADSII_xy} we see that sections ${\x,\ph=\text{constant}}$ 
can cross different number of horizons, depending on the value of ${\x}$, since
they can reach the infinity, given in this case by
\begin{equation}\label{scriII}
  \y=\coth\acp\;\x\commae
\end{equation}
before they cross the acceleration or cosmological horizons.
There are three different generic classes of sections ${\x,\ph=\text{constant}}$
labeled ${\sect{A}}$ (sections crossing all horizons), 
${\sect{B}}$ (sections which do not cross cosmological horizons),
and ${\sect{C}}$ (which cross only black hole horizons). Special limiting cases are 
${\x=\xc=\yc\tanh\acp}$ and ${\x=\xa=\ya\tanh\acp}$.
For each of these sections a different shape of conformal diagram is obtained as
can be found in Fig.~\ref{fig:CADSII_cd}.
For section ${\sect{A}}$ the domain II outside a black hole is connected
through the acceleration horizon to domains of type I which are connected
through other acceleration horizons to another domain II with another black hole.
The domain I is also connected through the cosmological horizon with 
two domains of type O. From these domains it possible to reach another domain I, and so on.

The spacetime thus seems to describe a universe which at one moment contains
a pair of black holes (domains III and IV) separated by the 
acceleration horizon (domains II and I), and at another moment does not contain any
black hole (domains I and O)---see Fig.~\ref{fig:CADSII_cd}a. 
However, the sections ${\sect{B}}$ do not contain domains O,
and sections ${\sect{C}}$ do not even contain  the domains I. How is it possible that
one spacetime is described by three qualitatively different diagrams?
And how is it possible that the spacetime with anti-de~Sitter 
asymptotic has a conformal diagram with conformal infinity which looks
spacelike as it occurs for sections ${\sect{B}}$ (see Fig.~\ref{fig:CADSII_cd}b)?

\begin{figure}[t]
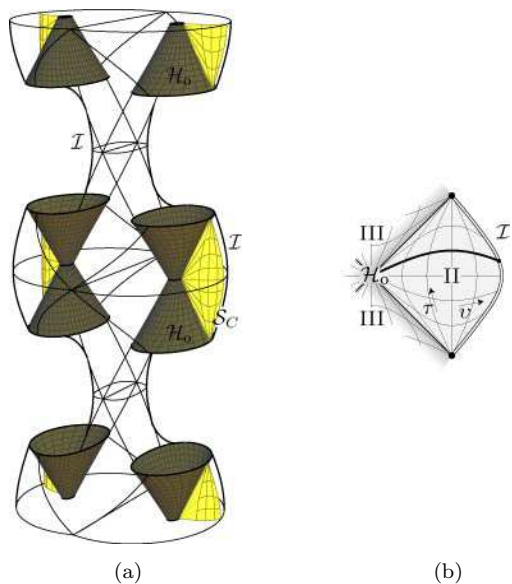

  \begin{minipage}[c]{110pt}\centering\includegraphics{\imgver fig_CAdSII_3DC}\end{minipage}
  \begin{minipage}[c]{110pt}\centering\includegraphics{\imgver fig_CAdSII_2DC}\end{minipage}\\[6pt]
  {\footnotesize\hspace*{10pt}(a)\hspace{110pt}(b)\hspace*{0pt}}
  \caption{\label{fig:CADSII_3DC}%
(a) Embedding of section ${\sect{C}}$
(cf.\ Figs.~\ref{fig:CADSII_xy} and \ref{fig:CADSII_cd}c) into a three-dimensional diagram. 
(b) The corresponding part of the two-dimensional conformal diagram of ${\sect{C}}$
(cf.\ the dark areas in Fig.~\ref{fig:CADSII_cd}c).  
}
\end{figure}

The answer can be given by drawing three-dimensional
diagram obtained by \vague{gluing} different sections of ${\x=\text{constant}}$
together. The inspiration how to do it can be obtained by a study of accelerated
static coordinates in empty anti-de~Sitter universe as will be done in Sec.~\ref{sc:AdS}.
There we will learn that coordinates ${\tau,\y,\x,\ph}$ (or ${\Tac,\Rac,\Thac,\Phac}$)
are sorts of bi-polar coordinates---coordinates with two poles centered on two black holes.
The coordinate ${\Rac}$ (respectively ${\y}$) is running through domain II 
from both black holes toward the acceleration horizon.
It plays the role of a radial coordinate in domain II, but it 
changes its meaning into a time coordinate above and below 
the acceleration horizon, in domains of type I.
It becomes again a space coordinate in domains O.
The angular coordinates ${\Thac,\Phac}$ (or ${\x,\ph}$) label
different directions connecting the two holes. With this insight we can draw 
the three-dimensional diagrams reflecting the global structure of the universe, see Fig.~\ref{fig:CADSII_3D}.
Embeddings of three typical surfaces ${\x,\ph=\text{constant}}$ into such a diagram
are shown in Figs.~\ref{fig:CADSII_3DA}--\ref{fig:CADSII_3DC}.
Here we can see an origin of different shapes of conformal diagrams.  

The global picture of the universe is thus the following: 
into an empty anti-de~Sitter-like universe (domains O and~I) 
enters through the infinity ${\scri}$\, a pair of black holes 
(domains III and IV). 
The holes are flying toward each other (domains II) with deceleration 
until they stop and fly back to the infinity where they leave the universe.
They are causally disconnected by the acceleration horizon. 
There follows a new phase without black holes (again, the domain I and O)
followed by a new phase with a pair of black holes. Different pairs
of black holes are separated by cosmological horizons.

Again, for purpose of the weak field limit it is convenient 
to use a visualization with squeezed black hole horizons
in Fig.~\ref{fig:CADSII_3D}b. In this representation, 
the infinity has a shape which one would expect for asymptotically
anti-de~Sitter universe. The deformation of the infinity
is related to the fact that we use coordinates 
centered around the black holes. Indeed, the black holes 
are drawn along straight lines in the vertical direction.
As we will see in the next section, 
such a deformation of the infinity is obtained even for 
an empty anti-de~Sitter universe if it is represented using
accelerated coordinates.

\section{Anti-de~Sitter universe in accelerated coordinates}
\label{sc:AdS}

The spacetime \eqref{KWmetric} reduces to the anti-de~Sitter 
universe for ${\mass=0,\,\charge=0}$. However, the limiting
metric is not the anti-de~Sitter metric in standard cosmological coordinates. 
Instead, it is the anti-de~Sitter metric in so-called accelerated 
coordinates which prefer certain accelerated observers. 
These observers are remnants of the black holes.
Investigating this form of the anti-de~Sitter metric is useful
for understanding of asymptotical structure of the $C$-metric universe,
and of the nature of the coordinate systems used.

The anti-de~Sitter metric can be written in cosmological spherical
coordinates ${\tE,\chi,\tht,\ph}$ as
\begin{equation}\label{AdSsph}
\mtrc_\AdS=\frac{\scl^2}{\cos^2\chi}
\Bigl(-\grad\tE^2+\grad\chi^2+\sin^2\chi\,\bigl(\grad\tht^2+\sin^2\tht\,\grad\ph^2\bigr)\Bigr)\period
\end{equation}
They can be also called conformally Einstein because they are the standard 
coordinates on the conformally related Einstein universe.
Another useful set of coordinates are
cosmological cylindrical coordinates ${\tE,\zC,\rC,\ph}$ which redefine
coordinates ${\chi}$ and ${\tht}$. Surfaces ${\tE,\rC=\text{constant}}$ represent
cylinders of constant distance from the axis, and surfaces ${\tE,\zC=\text{constant}}$
are planes orthogonal to the axis. They are related to spherical coordinates
by a rotation on the conformally related sphere of the Einstein universe by an angle ${\pi/2}$:
\begin{equation}\label{sphcylrel}
\begin{aligned}
  \cos\chi&=\cos\zC\cos\rC\commae\\
  \tan\tht&=\cot\zC\sin\rC\commae
\end{aligned}\quad
\begin{aligned}
  \sin\zC&=\sin\chi\cos\tht\commae\\
  \tan\rC&=\tan\chi\sin\tht\period
\end{aligned}
\end{equation}
The metric in the cylindrical coordinates reads
\begin{equation}\label{AdScyl}
\mtrc_\AdS\!=\!\frac{\scl^2}{\cos^2\!\zC\cos^2\!\rC}
\Bigl(\!-\grad\tE^2+\grad\zC^2+\cos^2\!\zC\,\bigl(\grad\rC^2+\sin^2\!\rC\,\grad\ph^2\bigr)\Bigr)\!\period
\end{equation}

The anti-de~Sitter universe admits four qualitatively different types
of Killing vectors representing time translations, boosts, null boosts,
and spatial rotations. Orbits of time translations
and boosts correspond to worldlines of observers with 
uniform acceleration. The limit of the $C$-metric is related
exactly to these observers. The cases ${\accl<1/\scl}$
and ${\accl>1/\scl}$ correspond to time translation and 
boost Killing vectors respectively; the case ${\accl=1/\scl}$
corresponds to a null boost Killing vector.

It is possible to introduce static coordinates associated with
the Killing vector that is at least partially timelike. In the case 
of the time translation Killing vector, both cosmological spherical
and cylindrical coordinates play the roles of such coordinates.
It is also possible to rescale the radial coordinate ${\chi}$ to obtain
metric in \vague{standard static} form. Namely, defining static coordinates
of type I
\begin{equation}\label{statI}
  \TI=\scl\,\tE\comma\RI=\scl\,\tan\chi\comma\ThI=\tht\comma\PhI=\ph
\end{equation}
we obtain
\begin{equation}\label{AdSstatI}
\begin{split}
\mtrc_\AdS &=
  -\Bigl(1+\frac{\RI^2}{\scl^2}\Bigr)\grad\TI^2
  +\Bigl(1+\frac{\RI^2}{\scl^2}\Bigr)^{\!-1}\grad\RI^2\\
  &\mspace{140mu}+\RI^2\bigl(\grad\ThI^2+\sin^2\!\ThI\,\grad\PhI^2\bigr)
  \period
\end{split}
\end{equation}

Static coordinates of type II are associated with the boost Killing vector 
and can be related to the cosmological cylindrical coordinates
\begin{equation}\label{statII}
\begin{gathered}
  \TII=\frac\scl2\log\abs{\frac{\sin\tE-\sin\zC}{\sin\tE+\sin\zC}}\comma
  \RII=\scl\frac{\cos\zC}{\cos\tE}\commae\\
  \ThII=\rC\comma\PhII=\ph\commae
\end{gathered}
\end{equation}
leading to the metric
\begin{equation}\label{AdSstatII}
\begin{split}
\mtrc_\AdS &= \frac{\scl^2}{\RII^2\cos^2\!\ThII}\Biggl[
  -\Bigl(1-\frac{\RII^2}{\scl^2}\Bigr)\grad\TII^2\\
  &+\Bigl(1-\frac{\RII^2}{\scl^2}\Bigr)^{\!-1}\grad\RII^2
  +\RII^2\bigl(\grad\ThII^2+\sin^2\!\ThII\,\grad\PhII^2\bigr)
  \Biggr]\period
\end{split}\raisetag{10ex}
\end{equation}

\begin{figure}
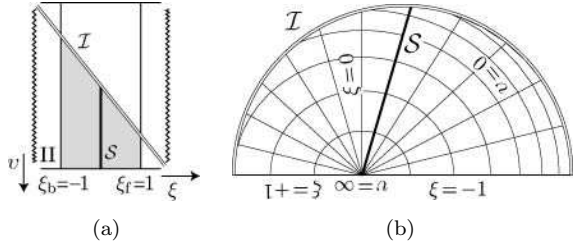

  \begin{minipage}[c]{80pt}\centering\includegraphics{\imgver fig_AdSI_xy}\end{minipage}
  \begin{minipage}[c]{140pt}\centering\includegraphics{\imgver fig_AdSI_2Dxy}\end{minipage}
  \\[6pt]
  {\footnotesize\hspace*{0pt}(a)\hspace{100pt}(b)\hspace*{30pt}}
  \caption{\label{fig:ADSI_xy}%
A shaded region in diagram (a) indicates allowed ranges of coordinates 
${\y\equiv\yI'=\scl/\RI'=\cot\chI'}$ and ${\x\equiv\xI'=}{-\cos\ThI'=-\cos\thI'}$.
The diagonal double-line corresponds to the infinity, 
vertical borders to the axis of symmetry and the bottom line to the origin 
${\chI'=0}$. The diagram (a) is an analogue of Fig.~\ref{fig:CADSI_xy}.
However, this diagram does not respect the angular meaning of the ${\x}$
coordinate. A more natural representation (b) of the shaded region
is obtained by shrinking the bottom line to a point, forming thus
a deformed semicircle.}
\end{figure}

In the case of the full $C$-metric we do not have to use a different 
notation for coordinates defined in the case ${\accl<1/\scl}$ and ${\accl>1/\scl}$,
because these two cases describe completely different spacetimes, and the
coordinates cannot be mixed. However, in the weak field limit
both cases describe one spacetime---anti-de~Sitter universe---and 
we have a whole set of coordinate systems, parametrized by acceleration,
living on this spacetime.
To avoid a confusion, in the next two subsections we add a prime and
subscript I (for ${\accl<1/\scl}$) or II (for ${\accl>1/\scl}$)
to all coordinates introduced in the previous sections.\footnote{%
We use the subscript to distinguish two qualitatively 
different cases (although, we still have a hidden parametrization 
of the coordinate systems by the acceleration), and the prime to 
indicate a nontrivial acceleration. Corresponding unprimed coordinates
refer to special values of the acceleration: $\accl=0$ in the case I, 
and ${\accl=1/\scl}$ in the case II. This notation is consistent with Ref.~\cite{Krtous:BIAS}.}
For example,
accelerated static coordinates ${\Tac,\Rac,\Thac,\Phac}$ will be renamed 
as ${\TI',\RI',\ThI',\PhI'}$ or ${\TII',\RII',\ThII',\PhII'}$
for small or large acceleration, respectively. 

Let us note that for both cases ${\accl\lessgtr1/\scl}$\,
in the weak field limit,
the metric function ${\G}$ reduces
to ${\G=1-\x^2}$ (see \eqref{CFGI} and \eqref{CFGII}).
By integrating \eqref{acdef} we then get ${\x=-\cos\Thac}$ 
and ${\G=\sin^2\Thac}$.

\subsection{${\accl<1/\scl}$}

\begin{figure}
  \includegraphics{\imgver fig_AdSI_cd}\\[9pt]
  {\footnotesize\hspace*{0pt}(a)\hspace{105pt}(b)\hspace*{0pt}}
  \caption{\label{fig:ADSI_cd}%
Conformal diagrams of the section ${\ThI',\PhI'=\text{constant}}$
(or, equivalently, ${\thI',\phI'=\text{constant}}$).
Diagram (a) is based on coordinates ${\uBI',\vBI'}$. 
Horizontal and vertical lines are given by coordinate lines 
${\tEI'=\text{constant}}$ and ${\chI'=\text{constant}}$,
since for ${\mass,\charge=0}$, definitions \eqref{ytort} and \eqref{uvBlock}
give ${\uBI'=\chI'+\tEI'}$ and ${\vBI'=\chI'-\tEI'}$.
The coordinate ${\chI'}$ is a radial coordinate; 
the left border of the diagram thus corresponds to the worldline
of an accelerated observer at the origin. The right double-line represents
conformal infinity ${\scri}$ (formed by limiting end points of spacelike and null geodetics).
The diagram should continue infinitely in the vertical direction.
(b) Compactified version of the same conformal diagram based
on the coordinates ${\uGI',\vGI'}$, related to ${\uBI',\vBI'}$
by Eq.~\eqref{uvGlob}. 
The whole spacetime is here squeezed into 
a compact region which beside the conformal infinity 
includes also point-like future and past 
infinities (limiting end points of timelike geodesics). 
This diagram is analogous to those in Fig.~\ref{fig:CADSI_cd}.
An exact position of ${\scri}$\ 
depends on an angular direction of the plane of the diagram
(i.e., on a value of coordinate ${\thI'}$) through
the relation ${\tan\chI'=-\scl\cot\aca/\cos\thI'}$ (cf.\ Eq.~\eqref{scriI}).
For ${\accl=0}$ these diagrams reduces to the standard conformal diagrams
based on the cosmological spherical coordinates.}
%\end{figure}
\bigskip\bigskip\bigskip
%\begin{figure}
  \includegraphics{\imgver fig_AdSI_3D}\\[9pt]
  {\footnotesize\hspace*{0pt}(a)\hspace{105pt}(b)\hspace*{10pt}}
  \caption{\label{fig:ADSI_3D}%
Three-dimensional schematical diagrams of anti-de~Sitter universe
obtained by rotation (varying angular coordinate ${\thI'}$) of 
two-dimensional diagrams from Fig.~\ref{fig:ADSI_cd}. 
The diagrams are centered on the worldline of a static
observer (thick line) which is accelerated with acceleration
${\accl<1/\scl}$. The horizontal section ${\tE=\text{constant}}$
corresponds to two copies of Fig.~\ref{fig:ADSI_xy}b
(one copy for ${\ph=0}$, another for ${\ph=\pi}$).}
\end{figure}

In the limit of vanishing mass and charge,  
the metric \eqref{Acmetric} with ${\Hac}$ given by Eq.~\eqref{HacI} takes 
the form\pagebreak[1]
\begin{equation}\label{AdSacI}
\begin{split}
\mtrc &=\frac{\scl^2}{\bigl(\scl\cos\aca\!+\!\RI'\sin\aca\cos\ThI'\bigr)^{2}}\Biggl[
  -\Bigl(1+\frac{\RI'^2}{\scl^2}\Bigr)\grad\TI'^2\\
  &+\Bigl(1+\frac{\RI'^2}{\scl^2}\Bigr)^{\!-1}\grad\RI'^2
  +\RI'^2\bigl(\grad\ThI'^2+\sin^2\!\ThI'\,\grad\PhI'^2\bigr)
  \Biggr]\period
\end{split}%\raisetag{10ex}
\end{equation}
The allowed ranges of coordinates 
can be read from Fig.~\ref{fig:ADSI_xy}.
For vanishing acceleration, ${\aca=0}$, the metric
becomes exactly of the form \eqref{AdSstatI}; i.e., $C$-metric
accelerated static coordinates become anti-de~Sitter static coordinates of type I.
For non-vanishing acceleration the form of the metric \eqref{AdSacI} differs from \eqref{AdSstatI}
by a scalar prefactor. However, we still claim that ${\mtrc=\mtrc_\AdS}$. 
The relation between coordinates ${\TI,\RI,\ThI,\PhI}$
and ${\TI',\RI',\ThI',\PhI'}$ is thus a coordinate conformal 
transformation of anti-de~Sitter space. It has a nice geometrical interpretation:
if we define accelerated spherical coordinates of type I,\; ${\tEI',\chI',\thI',\phI'}$, 
related to ${\TI',\RI',\ThI',\PhI'}$
analogously to definition \eqref{statI}, these coordinates differ from ${\tE,\chi,\tht,\ph}$
only by a rotation of the Einstein sphere in the 
direction of the axis ${\tht=\pi}$ by the angle ${\aca}$,\pagebreak[1]
\begin{equation}\label{statacI}
\begin{aligned}
  \tEI'&=\tE\commae&
  \cos\chI'&=\cos\aca\cos\chi-\sin\aca\sin\chi\cos\tht\commae\\
  \phI'&=\ph\commae&
  \cot\thI'&=\cos\aca\cot\tht+\sin\aca\cot\chi\sin^{\!-1}\tht\period
\end{aligned}
\end{equation}
Coordinates ${\tEI',\chI',\thI',\phI'}$ are thus a sort of spherical 
coordinates\footnote{%
They are spherical in the sense of conformally related Einstein universe.}
centered on the observer given by ${\chi=\aca}$, ${\tht=\pi}$. 
This observer remains eternally at a constant distance from the origin ${\chi=0}$,
and has a unique acceleration of magnitude ${\accl=\sin\aca}$
which compensates for the cosmological compression of anti-de~Sitter universe.
For more details see \cite{Krtous:BIAS}.

Two-dimensional conformal diagrams of ${\TI'\textdash\RI'}$ sections 
(i.e., of ${\tEI'\textdash\chI'}$ sections) 
can be found in Fig.~\ref{fig:ADSI_cd}.
Three-dimensional diagrams obtained by gluing together
two-dimensional sections with changing ${\ThI'}$ are
in Fig.~\ref{fig:ADSI_3D}. The diagram in Fig.~\ref{fig:ADSI_3D}b
is clearly the limiting case of Fig.~\ref{fig:CADSI_3Ddrop}.

\begin{figure}
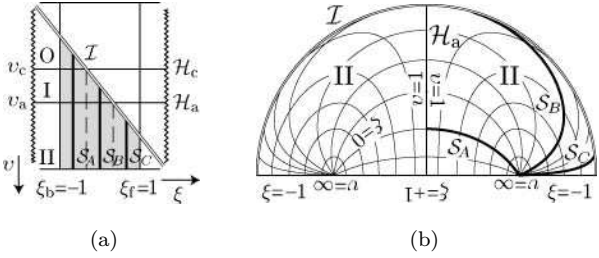

  \begin{minipage}[c]{80pt}\centering\includegraphics{\imgver fig_AdSII_xy}\end{minipage}
  \begin{minipage}[c]{160pt}\centering\includegraphics{\imgver fig_AdSII_2Dxy}\end{minipage}\\[9pt]
  {\footnotesize\hspace*{0pt}(a)\hspace{110pt}(b)\hspace*{40pt}}
  \caption{\label{fig:ADSII_xy}%
A shaded region in diagram (a) represents the allowed range of coordinates 
${\y\equiv\yII'=\scl/\RII'}$ and ${\x\equiv\xII'=-\cos\ThII'}$.
The notation is the same as in Fig.~\ref{fig:CADSII_xy},
except there are no black hole horizons, and the bottom line
does not represent a singularity but poles of the coordinates.
Diagram (a) does not respect the bi-polar nature
of coordinates ${\y}$ and ${\x}$. A more accurate 
picture of region II is drawn in diagram (b).
It depicts section ${\tE,\ph=0}$ through two spacetime domains of 
type II. Each of them contains one pole of the coordinate system.
Both domains are separated by an acceleration horizon.
The coordinate ${\y}$ decreases from ${\y=+\infty}$ at poles to ${\y=\ya}$
at the acceleration horizon, and the coordinate ${\x}$ labels 
different coordinate lines starting from the poles.}
\end{figure}

\subsection{${\accl>1/\scl}$}

\begin{figure}
  \includegraphics{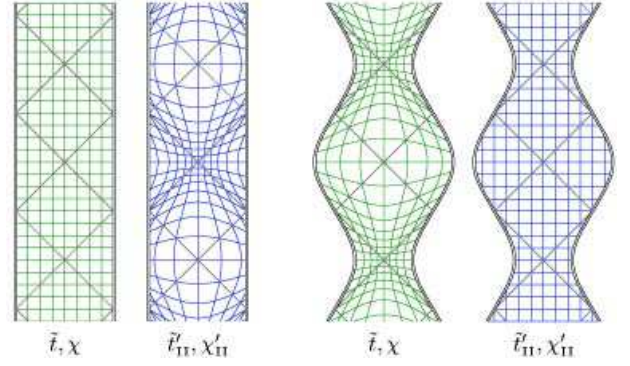}
  \caption{\label{fig:Sqeezing}%
Cosmological spherical coordinates ${\tE,\chi}$
and accelerated spherical coordinates ${\tEII',\chII'}$
drawn on a two-dimensional section of anti-de~Sitter universe.
The coordinate systems are related by 
the \vague{squeezing transformation} \eqref{sphaccrel}.
Left: Coordinate lines of both systems
drawn in such a way that lines
${\tE=\text{constant}}$ and ${\chi=\text{constant}}$
are horizontal and vertical, respectively. 
Right: A complementary representation with
vertical and horizontal lines given by coordinate system ${\tEII',\chII'}$.
The conformal infinity is given by ${\chi=\pi/2}$ and
is thus deformed in the squeezed diagram on the right.
}
\end{figure}

In this case,  
the metric \eqref{Acmetric} with ${\Hac}$ given by Eq.~\eqref{HacII} 
for vanishing mass and charge becomes\pagebreak[1]
\begin{equation}\label{AdSacII}
\begin{split}
\mtrc &=\frac{\scl^2}{\bigl(\scl\sh\acp\!+\!\RII'\ch\acp\cos\ThII'\bigr)^{2}}\Biggl[
  -\Bigl(1-\frac{\RII'^2}{\scl^2}\Bigr)\grad\TII'^2\\
  &+\Bigl(1-\frac{\RII'^2}{\scl^2}\Bigr)^{\!-1}\grad\RII'^2
  +\Rac^2\bigl(\grad\ThII'^2+\sin^2\!\ThII'\,\grad\PhII'^2\bigr)
  \Biggr]\period
\end{split}%\raisetag{10ex}
\end{equation}
The allowed range of coordinates ${\RII',\ThII'}$ 
can be read from Fig.~\ref{fig:ADSII_xy}.
For ${\acp=0}$ (i.e., in the limit ${\accl\to1/\scl}$) 
we get exactly the metric \eqref{AdSstatII}.
For nonzero ${\acp}$ both metrics \eqref{AdSacII} and 
\eqref{AdSstatII} have the same form up to a scalar prefactor. 
However, as in the previous case, it is possible to
find a transformation between ${\TII',\RII',\ThII',\PhII'}$
and ${\TII,\RII,\ThII,\PhII}$ such that ${\mtrc=\mtrc_\AdS}$.
First, we introduce accelerated spherical and cylindrical coordinates of 
type~II,\; ${\tEII',\chII',\thII',\phII'}$ and ${\tEII',\zCII',\rCII',\phII'}$, 
which are related to ${\TII',\RII',\ThII',\PhII'}$
as  ${\tE,\chi,\tht,\ph}$ and ${\tE,\zC,\rC,\ph}$ 
are related to ${\TII,\RII,\ThII,\PhII}$, i.e., by
the relations \eqref{sphcylrel} and \eqref{statII}.
Transformations between cosmological and accelerated coordinates then are
\begin{equation}\label{sphaccrel}
\begin{gathered}
\begin{aligned}
\cot\tEII' &= \frac{\mspace{16mu}\ch\acp\cos\tE-\sh\acp\cos\chi}{\sin\tE}\commae\\
\cot\chII' &= \frac{-\sh\acp\cos\tE+\ch\acp\cos\chi}{\sin\chi}\commae
\end{aligned}\\
\thII'=\tht\comma\phII'=\ph\period
\end{gathered}
\end{equation}
It is interesting, that these transformations leave angular coordinates untouched.
It means that they are a time dependent \emph{radial} \vague{sqeezing} of anti-de~Sitter universe;
see Fig.~\ref{fig:Sqeezing}. 

\begin{figure}
  \includegraphics{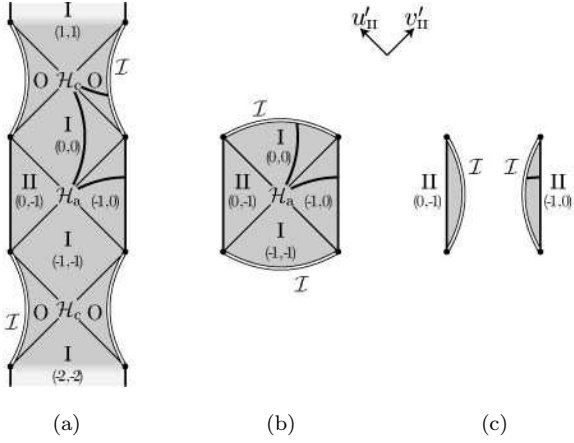}\\[6pt]
  {\footnotesize\hspace*{0pt}(a)\hspace{70pt}(b)\hspace{70pt}(c)\hspace*{0pt}}
  \caption{\label{fig:ADSII_cd}%
The conformal diagrams of sections ${\xII',\phII'=\text{constant}}$ 
(or, equivalently, ${\rCII',\phII'=\text{constant}}$)
for different values of ${\x\equiv\xII'}$. The diagrams are based
on null coordinates ${\uGII',\vGII'}$. They are spanned between 
two poles which corresponds to observers with uniform acceleration ${\accl>1/\scl}$
(straight vertical lines on the border of the diagrams,
cf.\ also Fig.~\ref{fig:ADSII_3D} for the three-dimensional 
localization of the poles).
Three different shapes of the diagrams correspond to 
qualitatively different possibilities of how sections 
${\xII',\phII'=\text{constant}}$ are embedded into the anti-de~Sitter universe.
They correspond to the three sections ${\x=\text{constant}}$ indicated in Fig.~\ref{fig:ADSII_xy}.
Diagonal lines represent acceleration and cosmological horizons.
The acceleration horizon causally separates both poles. It is formed by future
light cones of points where the poles enter the anti-de~Sitter universe.
The cosmological horizon is formed by future light cones of points
where the poles leave the universe. The thick line is
an example of ${\y\equiv\yII'=\text{constant}}$
section---it corresponds to the diagram in Fig.~\ref{fig:ADSI_xy}.
Gluing together diagram (a) for ${\xII'=-1}$ (the axis 
${\rCII'=\ThII'=0}$ between poles)
with diagram (c) for ${\xII'=+1}$ (the axis ${\rCII'=\ThII'=\pi}$) 
gives the history
of the whole axis of symmetry. It is the same section as that
depicted in Fig.~\ref{fig:Sqeezing}. 
}
\end{figure}

Surprisingly, if we compose all 
partial transformations between ${\TII',\RII',\ThII',\PhII'}$
and ${\TII,\RII,\ThII,\PhII}$ together, the resulting transformation 
is such that ${\TII'=\TII}$ and ${\PhII'=\PhII}$, see Ref.~\cite{Krtous:BIAS}---time
surfaces of both the static coordinates of type II 
and of the accelerated static coordinates are the same.
 
Now, let us study global null coordinates ${\uGII,\vGII}$ related
to the static coordinates of type II\; ${\TII,\RII}$ by the relations
\eqref{uvBlock} and \eqref{uvGlob}. With vanishing mass and charge 
(and setting ${\GNcoef=1/2}$ in \eqref{uvGlob})
these definitions give
\begin{equation}\label{uvGcylrel}
\begin{aligned}
  \uGII&=\tE-\zC\commae& \ThII&=\rC\commae\\
  \vGII&=\tE+\zC\commae& \PhII&=\ph
\end{aligned}
\end{equation}
Horizontal and vertical lines of the conformal diagram based on ${\uGII,\vGII}$ 
are thus coordinate lines ${\tE=\text{constant}}$ and ${\zC=\text{constant}}$.
The surface of this conformal diagram, i.e., the surface ${\rC,\ph=\text{constant}}$, is
a history of a line with a constant distance from the axis of symmetry. 
All such lines have common limiting end points ${\zC=\pm\pi/2}$ located at the infinity 
of the anti-de~Sitter universe. We will call them poles of the cylindrical coordinates.\footnote{
Lines of constant distance from the axis are not geodesics (except the axis itself)
in sense of the Lobachevsky geometry of the spatial section ${\tE=\text{constant}}$.
However, in the conformally related spherical geometry of the spatial section of Einstein universe,
these lines are meridians with common poles. These two poles lie on the 
boundary of the hemisphere which corresponds to the Lobachevsky plane, i.e., at its infinity.} 

\begin{figure}
  \includegraphics{\imgver fig_AdSII_3D}\\[9pt]
  {\footnotesize\hspace*{0pt}(a)\hspace{113pt}(b)\hspace*{7pt}}
  \caption{\label{fig:ADSII_3D}%
Three-dimensional representations of anti-de~Sitter universe
based (a) on cosmological coordinates ${\tE,\chi,\tht,\ph}$ 
and (b) on \vague{squeezed} accelerated coordinates ${\tEII',\chII',\thII',\phII'}$.
They can be obtained by a rotation of the corresponding diagrams from Fig.~\ref{fig:Sqeezing}.
Alternatively, they can be constructed by gluing together two-dimensional diagrams
from Fig.~\ref{fig:ADSII_cd}. These are spanned between worldlines of 
poles moving with the uniform acceleration ${\accl=1/\scl}$ along the axis. 
Worldlines of the poles are indicated in the diagrams by thick lines. 
A pair of the poles enter anti-de~Sitter universe through
the conformal infinity, they approach each other and then return
back to the infinity---all this in a finite cosmological time ${\Delta\tE=\pi}$.
After a stage without poles, a new pair of poles enters the universe, and so on.
The diagram (b) is clearly the limit of Fig.~\ref{fig:CADSII_3D}b
in which the black holes are shrunk to to the accelerated particles located at the poles.   
}
\end{figure}

The conformal diagrams constructed in Sec.~\ref{sc:CAdSII} are based on coordinates
${\uGII',\vGII'}$, i.e., on an \vague{accelerated} version of ${\uG,\vG}$ discussed in the previous paragraph.
For ${\mass,\charge=0}$, these diagrams are depicted in Fig.~\ref{fig:ADSII_cd}.
Different sections ${\rCII',\phII'=\text{constant}}$ again correspond
to histories of curves which end at common poles ${\zCII'=\pm\pi/2}$.
However, the infinity of the anti-de~Sitter universe 
in accelerated cylindrical coordinates is given by (cf.~\eqref{scriII})
\begin{equation}\label{scriIIAdS}
  \cos\zCII'\cos\rCII'=-\tanh\acp\cos\tEII'\period
\end{equation}
The poles thus, in general, do not lie at the infinity.
Coordinates ${\tEII',\zCII',\rCII',\phII'}$ are sort of
\vague{bi-polar coordinates} with poles which correspond
to the observers with acceleration ${\accl>1/\scl}$; 
see Fig.~\ref{fig:ADSI_xy}b. 
These observers, however, do not remain in anti-de~Sitter universe 
eternally. Their histories periodically enter and leave the spacetime
as shown in Fig.~\ref{fig:ADSII_3D}.
The section ${\rCII',\phII'=\text{constant}}$, spanned between 
the poles, intersect anti-de~Sitter universe in various
ways, depending on a value of ${\rCII'}$. 
Different intersections lead to qualitatively different 
conformal diagrams in Fig.~\ref{fig:ADSII_cd}.
This is in the agreement with analogous discussion
in Sec.~\ref{sc:CAdSII}.

\begin{acknowledgments}
The work was supported in part by program 360/2005 
of Ministry of Education, Youth and Sports of Czech Republic.
The author is grateful for kind hospitality at the Division of Geometric Analysis and Gravitation,
Albert Einstein Institute, Golm, Germany, and the Department 
of Physics, University of Alberta, Edmonton, Canada
where this work was partially done. 
He also thanks Don~N.~Page for reading the manuscript. 
% and for adding articles to the article.
\end{acknowledgments}

\newpage
%\bibliography{D:/odborne/library/TeX/bib/references,D:/odborne/library/TeX/bib/references_priv}
%\end{document}

\end{document}